\documentclass{aastex62}

\newcommand{\rep}[1]{{ #1}}

\received{\today}
\revised{\today}
\accepted{\today}
\submitjournal{AJ}

\shorttitle{DRAG{\sc races}}
\shortauthors{Chen\'e et al.}
\begin{document}

\title{DRAG{\sc races}: A pipeline for the GRACES high-resolution spectrograph at Gemini\footnote{First released on September, 20th, 2016}}

\correspondingauthor{Andr\'e-Nicolas Chen\'e}
\email{andrenicolas.chene@gmail.com}

\author[0000-0002-1115-6559]{Andr\'e-Nicolas Chen\'e}
\affil{Gemini Observatory/NSF’s NOIRLab, 670 N. A`ohoku Place, Hilo, Hawai`i, 96720, USA}

\author{Shunyuan Mao}
\affiliation{Department of Physics and Astronomy, 
University of Victoria,\\
PO Box 1700 STN CSC,
Victoria, BC  V8W 2Y2,
Canada}

\author[0000-0001-9589-3793]{Michael Lundquist}
\affiliation{University of Arizona, Steward Observatory, 933 N Cherry Ave, Tucson, AZ 85719, USA}

\collaboration{(Gemini collaboration)}

\author[0000-0002-5084-168X]{Eder Martioli}
\affiliation{Institut d'Astrophysique de Paris, UMR7095 CNRS, Universit\'e Pierre \& Marie Curie, \\
98bis boulevard Arago, 75014 Paris, France}
\affiliation{Laborat\'{o}rio Nacional de Astrof\'{i}sica, Rua Estados Unidos 154, \\
37504-364, Itajub\'{a} - MG, Brazil}
\altaffiliation{CFHT Corporation, 65-1238 Mamalahoa Hwy, Kamuela, HI 96743, USA}

%\author{Lison Malo}
%\affiliation{Institute for Research on Exoplanets, \\
%D\'epartement de physique, Universit\'e de Montr\'eal, \\
%CP 6128, Succursale Centre-Ville, Montr\'eal, Quebec H3C 3J7, Canada}
%\altaffiliation{CFHT Corporation,
%65-1238 Mamalahoa Hwy, Kamuela, HI 96743, USA}

\collaboration{(OPERA collaboration)}
\author[0000-0002-3936-9628]{Jeffrey L. Carlin}
\affiliation{Vera C. Rubin Observatory, Legacy Survey of Space and Time (LSST),\\ 950 North Cherry Avenue, Tucson, AZ 85719, USA }

%\author{Allyson Sheffield}
%\affiliation{Department of Natural Science, City University of New York ,\\ LaGuardia Community College, Long Island City, NY 11101, USA}

\collaboration{(Usability testing)}



\begin{abstract}

This paper describes the software DRAG{\sc races} (Data Reduction and Analysis for GRACES), which is a pipeline reducing spectra from GRACES (Gemini Remote Access to the CFHT ESPaDOnS Spectrograph) at the Gemini North Telescope. The code is written in the IDL language. It is designed to find all the GRACES frames in a given directory, automatically determine the list of bias, flat, arc and science frames, and perform the whole reduction and extraction within a few minutes. We compare the output from DRAG{\sc races} with that of OPERA, a pipeline developed at CFHT that also can extract GRACES spectra. Both pipelines were developed completely independently, yet they give very similar extracted spectra. They both have their advantages and disadvantages\rep{. For instance, DRAG{\sc races} is more straightforward and easy to use and is less likely to be derailed by a parameter that needs to be tweaked, while OPERA offers a more careful extraction that can be significantly superior when the highest resolution is required and when the signal-to-noise ratio is low. One} should compare both before deciding which one to use for their science. Yet, both pipelines deliver a fairly comparable resolution power \rep{($R\sim 52.8$k and 36.6k for DRAG{\sc races} and $R\sim 58$k and 40k for OPERA in high and low-resolution spectral modes, respectively)}, wavelength solution and signal-to-noise ratio per resolution element.

\end{abstract}



\section{Introduction} \label{sec:intro}

Gemini Remote Access to CFHT ESPaDOnS Spectrograph (GRACES) is the high-resolution spectroscopy capability available at the Gemini North facility \citep{Ch14}. It is the result of a cooperation between the Canada-France-Hawaii Telescope (CFHT), Gemini, and NRC-HAA (Canada). It combines the large collecting area of the Gemini North telescope with the high resolving power and high efficiency of the ESPaDOnS spectrograph at CFHT, to deliver high resolution spectroscopy across the optical region. This is achieved through a 270 m fiber optic feed from the Gemini North telescope to ESPaDOnS. The on-sky diameter of the fiber is 1.2 arcsec.
GRACES achieves a maximum resolution power of $R \sim 60,000$ between 400 and 1,000 nm, with throughput redward of 600 nm higher than 10\%, comparable to those of currently available high-resolution spectrographs in 8-10 m class telescopes. The Gemini Multi-Object Spectrograph (GMOS-N) holds the fiber injector on the Gemini side, and is used for acquiring targets. The GMOS-N On-Instrument Wave Front Sensor (OIWFS) is primarily used for guiding, but the Peripheral WFSs could also be chosen under certain circumstances. On the CFHT side of the fiber, GRACES has its own bench with its own image slicer inside of the ESPaDOnS enclosure that feeds the light into the spectrograph.

GRACES has been offered at Gemini since the 2015B semester, and is planned to remain in operation until at least 2022. It is a visiting instrument, which means that it is not fully integrated into the Gemini system like a facility instrument would be. Among other things, Gemini does not support any data reduction software for GRACES. It depends on the open source pipeline OPERA, developed at CFHT \citep{Ma12}, which is run at the end of each of the GRACES observing runs. 

This paper describes the basic functionality of an alternative GRACES data reduction pipeline DRAG{\sc races} (for Data Reduction and Analysis for GRACES)\rep{, originally developed as a safety back-up during commissioning in case OPERA would not work with the GRACES data as expected,} contained in the v1.3.1 release. As with most software, DRAG{\sc races}' development is ongoing and features are being added regularly. DRAG{\sc races} is an open-source code, written in the IDL\footnote{Interactive Data Language} language, that is being developed under the git version-control system on GitHub:\\ \url{https://github.com/AndreNicolasChene/DRAGRACES/releases/tag/v1.3.1}. The code uses the IDL Astronomy Library \citep[astrolib,][]{La93}, as well as the imdisp.pro function for figures displaying 2D images. Please note that contrary to what the affiliation of the first authors would suggest, the Gemini observatory is not involved with DRAG{\sc races} development and maintenance.

\section{About GRACES} \label{sec:overview}

\subsection{Instrument's brief description}
GRACES consists of three components: 1) an injection module sending the light from the Gemini telescope into the GRACES fibers, 2) two 270m-long GRACES fibers and 3) a receiver unit responsible for injecting the light from the fibers into the ESPaDOnS spectrograph at the CFHT. A complete description of the GRACES components is presented by \citet{An12}. The receiver unit contains a bench holding the optics, an image slicer, a dekker blocking “unwanted” light from the slicer, a shutter and a pickoff mirror sending the light to the spectrograph. The bench can rotate to switch between two slicing modes. One mode slices the image of the two fibers into two parts each, giving a lower spectral resolution mode (R$\sim 40k$) called ``two-fiber spectral mode''. That mode is used when the sky is observed simultaneously with the target. The other mode slices the image of the one fiber in four, giving a higher spectral resolution mode (R$\sim 60k$) called ``one-fiber spectral mode''.

Because spectra are obtained through most of the same optics and recorded on the same detector as the ESPaDOnS spectrograph, the data format is very comparable to those of ESPaDOnS  at the CFHT \citep{Ma03}. The main differences are the position of the traces on the detector, the shape of the slices, and the telescope information (in the standard Gemini format) in the headers (see the next Section). 
%\textcolor{red}{ML: A sentence or two here on the differences between the data formats e.g. headers include telescope information in the standard Gemini format.  AN, is there anything that's different regarding the columns in the data compared to CFHT that should be addressed?}

\subsection{Data description and access}
\begin{figure}[ht!]
\plotone{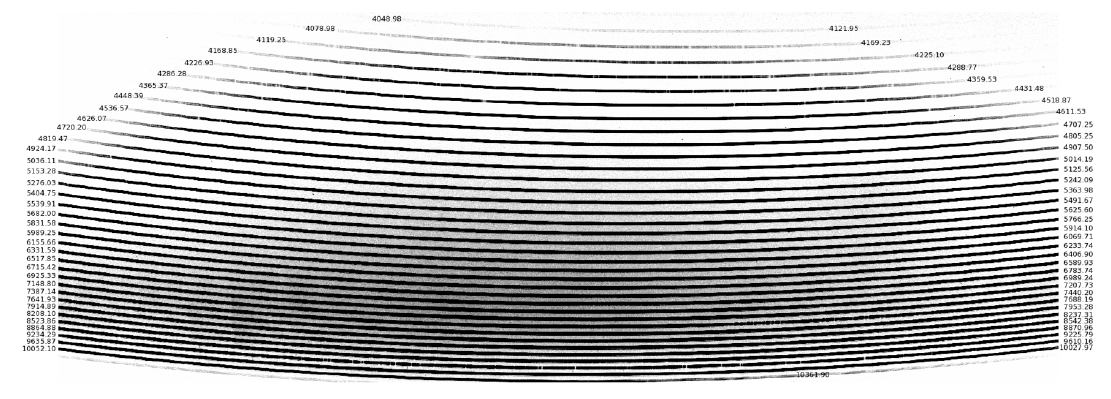}
\caption{This figure shows the raw 2D spectrum of the A3 star HIP 57258. It also is the GRACES first light! On each side are written the starting and ending wavelength for each of the orders. \rep{A 90$^\circ$ rotation was applied to the frame, so the rows are are now vertical and column horizontal.}\label{fig:couverture}}
\end{figure}

GRACES raw and reduced data can be accessed from the Gemini Observatory Archive (GOA) search form\footnote{\url{https://archive.gemini.edu/searchform}}. Each frame is recorded in a single extension fits file. An example of a stellar spectrum is presented in Figure\,\ref{fig:couverture}. Some scattered light is affecting the data, peaking somewhere around 7800\,\AA. Light contamination correction is covered in Section\,\ref{subsubsec:illum}.

The GRACES data are similar, but not identical to those of ESPaDOnS. First of all, GRACES data have merely any signal in bluer orders than the 56$^{\rm th}$. Also, the orders are not on the same pixels on the chip. At the central row of the detector, the difference in pixel of the center of the orders between the ESPaDOnS and GRACES spectra can be fairly well described with the parabola $40+0.16 x+0.032 x^2$, though this information is not used in the pipeline. The width of the orders are the same for the two instruments, but the shape is not identical. One can compare the shape from ESPaDOnS shown in \citep{Ma12} with the shape GRACES in \citep{Ch14}. Finally, the header for GRACES follows the Gemini standards\footnote{\url{http://www.gemini.edu/observing/phase-iii}} and the header for ESPaDOnS follow the CFHT standards.

The required calibration frames are 5 flat-field (quartz halogen lamp) exposures, 3 arc (Th-Ar lamp) exposures and 10 bias exposures. Calibrations are taken in the evening before and in the morning after the science observations were taken. There is a maximum of one change of spectral mode per night. If the spectral mode stayed the same all night long, the calibrations that were taken the closest in time with the science data should be used. If the spectral mode did change during the night, the calibrations with the mode corresponding to that of the science data must be used. \rep{Note that there are strict rules as part of Gemini's routine operation to guarantee that all the calibrations are always taken in the right modes for every night. Would any calibration go missing due to a mistake or unforeseen circumstances, all the science data of the night would not pass the quality assessment and the corresponding observations would be put back to the queue to be rescheduled.}

\subsection{Motivation} \label{sec:motivation}
GRACES was a quick and low-cost option to allow high-resolution spectroscopy to the Gemini Observatory. It of course takes advantage of the exceptional efficiency of the ESPaDOnS spectrograph, but also of the unprecedented quality of the 270m long fiber between the telescope and the instrument. GRACES is also a demonstrator of the possibility to leverage the use of a shared instrument between more than one observatory without having to move the instrument.

DRAG{\sc races} was first developed as a quick look tool for nighttime operations and was later expanded as a full pipeline. It was also developed as a back-up solution before it was known whether OPERA would be usable to extract GRACES data. Since OPERA ended up being successful in doing so, the Gemini Observatory delivers extracted data obtained using the OPERA pipeline \citep{Ma12} and DRAG{\sc races} became more of an alternative for those who would prefer using it \citep[e.g.,][]{Ca18}. But more importantly,{\rep DRAG{\sc races} has the advantage of being under development still, as opposed to OPERA for which all development have been discontinued. This means that DRAG{\sc races} can be continue to improve and evolve based on the users' feedback.  Since neither OPERA nor DRAG{\sc races} have been extensively tested using a wide range of data type and scenarios, comparing the pipelines' output is a safe way to proceed, and a potential way to improve DRAG{\sc races}' performance. DRAG{\sc races} may eventually be adopted as the default pipeline at Gemini, not only improve the analysis of future projects but also of the archival data.}

%\textcolor{red}{ML:Changed around some of the sentence structure.  Perhaps change `It was also developed as a back-up solution when it was still unknown if OPERA would be usable to extract GRACES data in time before the instrument was offered to the community.' to `It was also developed as a back-up solution before it was known whether OPERA would be usable to extract GRACES data.'}

%\subsection{}

\section{The pipeline} \label{sec:pipeline}

\begin{figure}[ht!]
\plotone{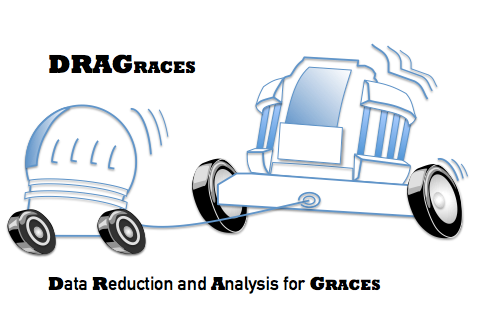}
\caption{DRAG{\sc races} logo features both the CFHT and Gemini North domes, linked together by the GRACES fiber. The CFHT dome is pictured as the driving force, illustrating the central role that the ESPaDOnS spectrograph and the CFHT support team play in GRACES. Drag races are also popular events on the Big Island of Hawai`i.\label{fig:logo}}
\end{figure}

The pipeline will find and process all the files it can find in a given directory. It can identify the different readmodes and spectral modes, and process them separately. In summary, the steps are:
\begin{itemize}
\item Sorting the fits files based on observation type \rep{(bias, flat, arc, object) as defined in the OBSTYPE header keyword}
\item Creating the master bias, master ThAr and master flatfield frames
\item Finding the trace of the orders on the master flatfield frame
\item Determining the slit tilt
\item Normalizing the flatfield in 2D
\item Identifying the arc lines and calculating the wavelength solution in 1D
\item Correcting for background illumination in 2D
\item Extracting the science spectra
\item Wavelength calibrating the science spectra
\end{itemize}

\subsection{Accessing the source code}

The pipeline is open source, version controlled, easy to use, and available to the community.\footnote{The source code and necessary files are accessible through GitHub: \\ \url{https://github.com/AndreNicolasChene/DRAGRACES}}

%\textcolor{red}{ML: Maybe to make this section a bit longer add text about DRAGraces being open source, version controlled, easy to use, and available to the community.  Then give the url as a footnote?}

\subsection{Assumptions}

\begin{itemize}
\item All the necessary files are in the same directory.
\item The spectrograph is stable enough, so the trace of the orders will always have the same width and the same shape.
\item The spectrograph is stable enough, so the orders will not move significantly on the detector within $\sim$10 hours of observations.
\item The spectrograph is stable enough, so the orders will never move by more than a few pixels with respect where they were during the 2015 commissioning.
\end{itemize}

These assumptions are supported by many years of calibration data. The only occasion where the position of the traces shifted on the detector was after several components needed to be replaced due to a lightning strike in August 2018. DRAG{\sc races} was updated to use the right hard-coded values in pixels where to search for the traces depending on if the data are from before or after the lightning event. 

%\textcolor{red}{ML: Do we need justify that we think the spectrograph is stable enough to satisfactorily make these assumptions?}

\subsection{List creation}

DRAG{\sc races} automatically creates lists of files. It begins by looking into the raw data directory collecting all of the FITS files with a filename that start with an 'N', followed by 8 digits (for the UT date), then the letter 'G', and 4 digits (for the file number), e.g., N20170611G0005.fits, which is the standard Gemini GRACES format.

All the frames are sorted based on the header keyword 'OBSTYPE', which can either be 'BIAS', 'FLAT', 'ARC' or 'OBJECT'. The bias frames are divided into slow ('READNOISE' = 2.9~e$^-$) and normal ('READNOISE' = 4.2~e$^-$) readmode. The other frames are divided into one-fiber ('GSLIPOS' = 'FOURSLICE') and two-fiber ('GSLIPOS' = 'TWOSLICE') spectral mode. Of course, the lists of bias, arcs or flats from the user is used instead of the automated ones, when provided.

If the data for more than one night are in the same directory, it is recommended to reduce each of the nights using the UT dates option. Otherwise, all the calibration files present will be combined and applied to all the object frames.

Sometimes, the calibrations obtained in the evening are in the same mode as the ones obtained in the morning. In such case, it is recommended to use only the set that is the closest in time to the spectra to process.

If the user wants to impose lists of bias, flat and/or arc frames, the filenames of the lists can be entered as an optional input (see dg.pro help). The lists have to be in the ASCII format with the fits files listed as a single column (one list per calibration type).

\subsection{The master calibration frames}
\noindent \texttt{Function: overs\_corr}\\ 
\indent Bias, arc, and flat frames are overscan corrected. The bias frames are combined first using a simple average. The resulting bias master frame is used to correct each individual arc and flat frames, before they are themselves combined into their respective master frames. Note that there are no cosmic ray rejection methods implemented at the moment. A visual inspection of the calibration frames is recommended before including them for the reduction. 

\subsection{Finding the orders}
\begin{figure}[ht!]
\plottwo{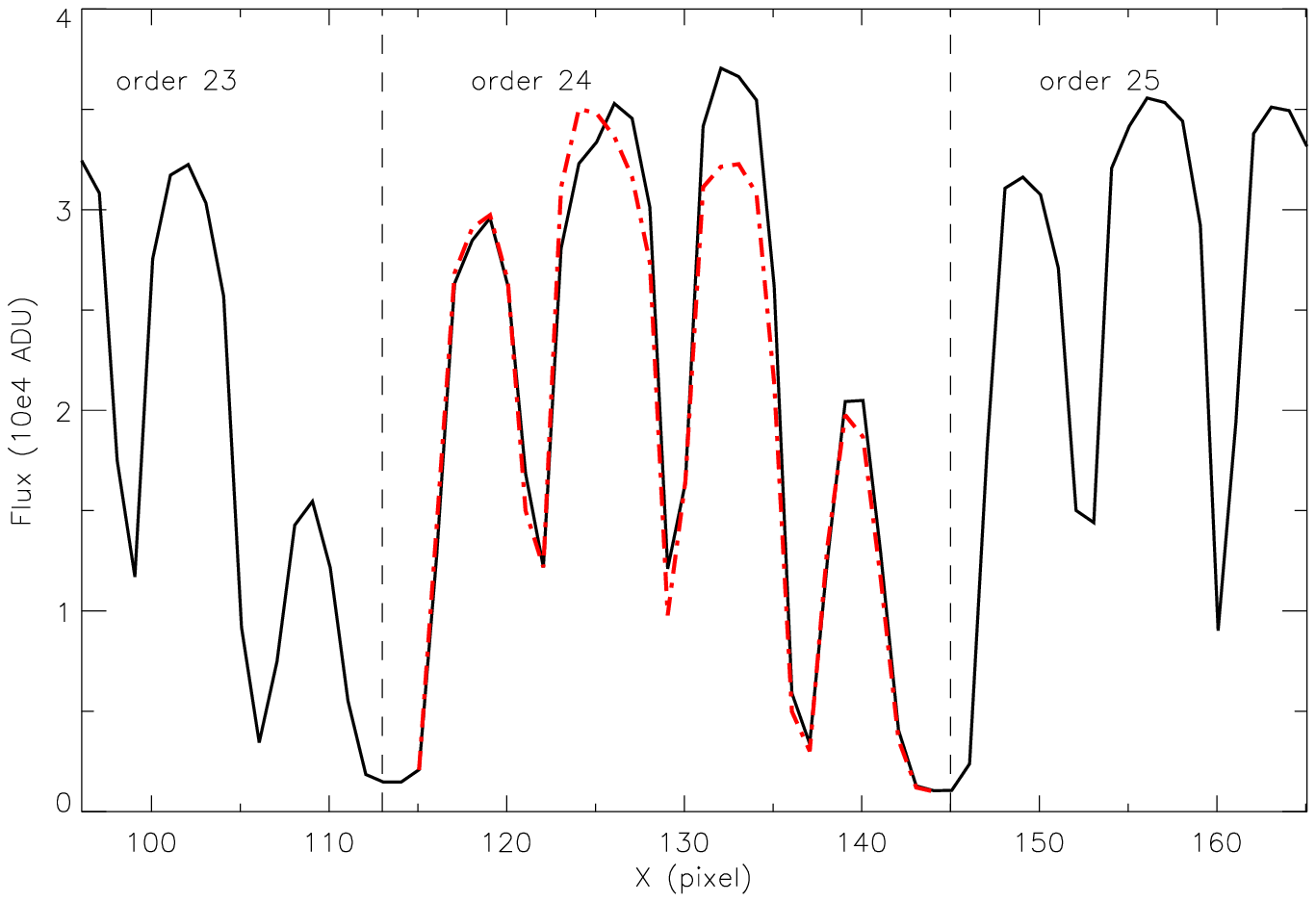}{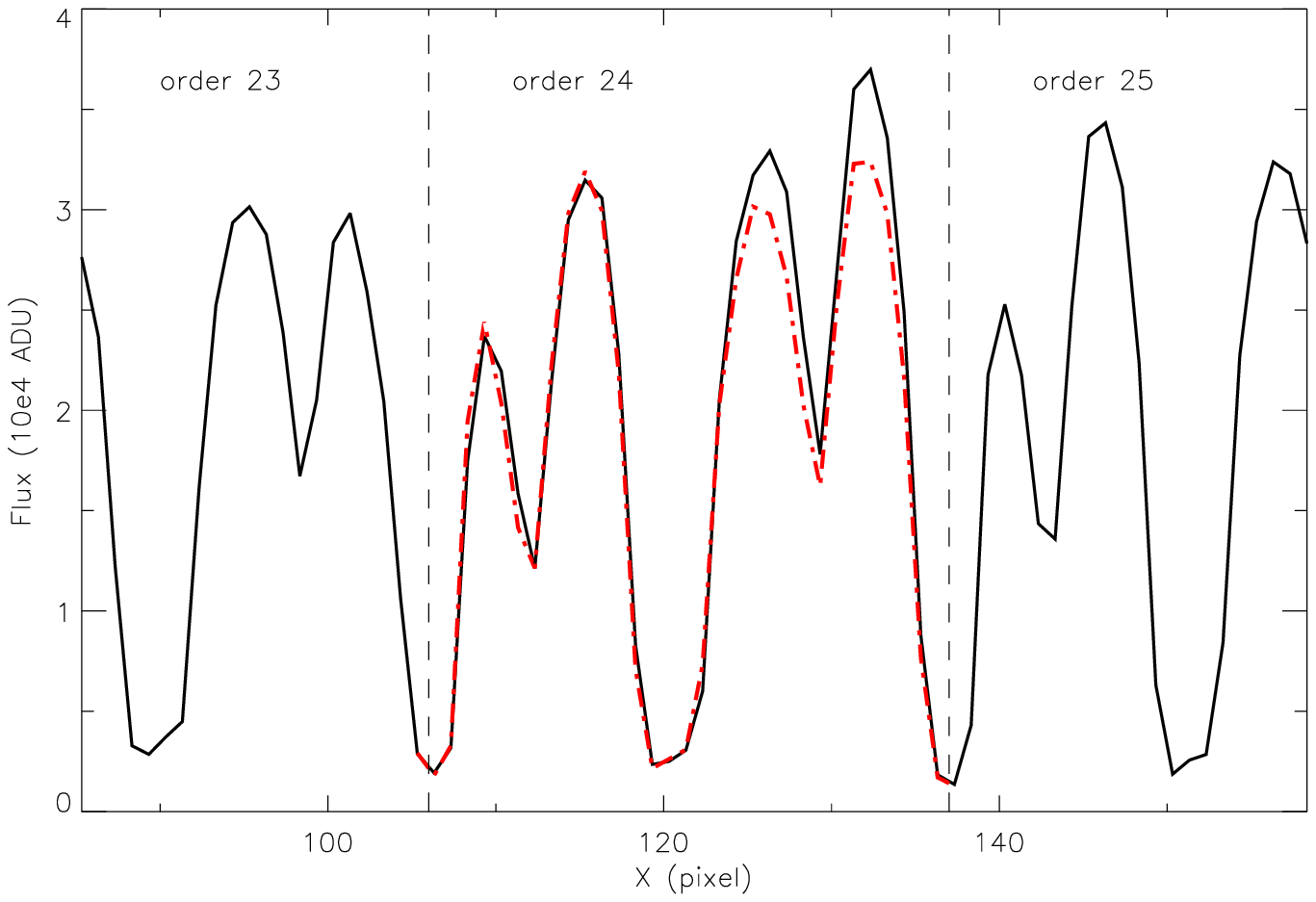}
\caption{Cut of the flatfield image through the \rep{row} 2004, centered on the order 24 for the one-fiber spectral mode (left panel, observe on 18 Jun 2017) and the two-fiber spectral mode (right panel, observed on 11 Jun 2017). The vertical dashed lines mark the limits of the order 24, and the red, dash-dotted line is the order's profile at \rep{row} 2304 (i.e., the middle of the detector) shifted to center on the selected trace. \label{fig:trace}}
\end{figure}

\noindent \texttt{Function: find\_trace} \\
\indent The order search starts in the middle of the detector, i.e. on the detector's \rep{row} 2304, using the master flatfield frame. At that \rep{row}, it is assumed that the center of the orders are near values that are hard coded in the pipeline. The level of stability of the instrument allows the assumption that those values are within a few pixels from the current centroids. 
The current centroids are determined using a cross-correlation technique to compare each order's profile (as observed on \rep{row} 2304, i.e., the middle of the detector) with a model, which is assumed to be a sum of 4 Gaussian functions with a sigma of 2 pixels and separated by a fixed number of pixels\rep{(see an example on Figure~\ref{fig:trace}).
Determining those centroids for each order within a one pixel accuracy is sufficient, since the goal is to isolate the profile of the trace at the middle \rep{row} of the detector. For each order, the profile at the row 2304 replaces the 4 Gaussian model in the cross-correlation that determines the centroid of the trace at every row}.
The measured centroids along the detector as a function of the X pixels (vertical) are fitted with a polynomial of order 4, and the coefficients are saved in a file to use later during the extraction. 

%\textcolor{red}{ML:Slight changes to the sentence on `the centroid for a given order is determined to be in the middle of the detector'}

\subsection{Determining the slit tilt}\label{subsec:tilt}
\noindent \texttt{Functions: reduce, extract, find\_lines} \\
\indent The slit tilt is introduced by the instrument, with the intention to prevent any overlap of the redder orders on the chip. While this is practical for optical considerations, it causes the need for an extra step before moving on with the reduction. 

To determine the slit tilt, the pipeline measures the slope of 10s of identified ThAr spectral lines per order on the arc master frame along each order.

\subsubsection{First ``crude'' wavelength solution and line identification}

The code contains a first estimate of the wavelength solution for each order. It was originally obtained using the function {\it IDENTIFY} from IRAF\footnote{IRAF is distributed by the National Optical Astronomy Observatory, which is operated by the Association of Universities for Research in Astronomy (AURA) under a cooperative agreement with the National Science Foundation.} on an extracted ThAr frame observed during GRACES commissioning. At the first pass, this estimated solution is applied to a 1-D arc master frame, extracted using a simple sum of all the columns in each order. The ``crude'' (i.e., not exact, but close) wavelength solution is assumed to be a shift of a few pixels from the initial one.

\begin{figure}[ht!]
\epsscale{0.55}
\plotone{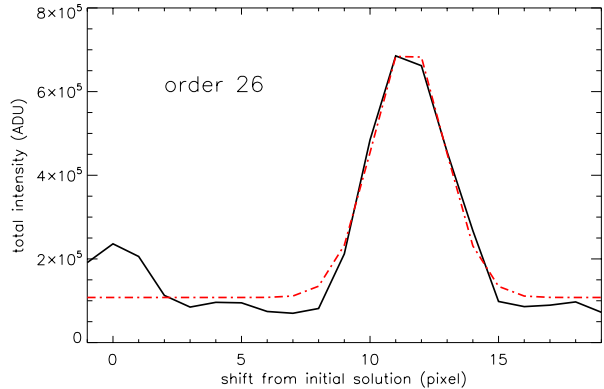}
\caption{Convolution of the order 26 of the observed ThAr with a comb function that consists of delta peaks with an amplitude of 1 at each wavelength where a spectra line is expected. The result is a weighted mean line profile. The red, dash-dotted line is the result of the Gaussian fit. Note that the line is wide due to the smearing caused by the slit tilt. \label{fig:ThArmean}}
\end{figure}

To determine and apply the shift for each order, the pipeline:
\begin{itemize}
\item creates a comb function that has a delta peak with an amplitude of one at each wavelength where a line is expected, following the ThAr line list given by \citet{atlas},
\item masks the saturated lines on the ThAr spectrum and all the pixels affected next to them,
\item convolves the observed ThAr spectrum with the comb function (the result is a single weighted mean profile, see Figure~\ref{fig:ThArmean}),
\item fits a Gaussian function on the resulting profile to determine the centroid (see Figure~\ref{fig:ThArmean}), and
\item shifts the first estimate of the wavelength solution by -1 times the value of the centroid determined by the fit.
\end{itemize}

\subsubsection{Measure of the tilt}

\begin{figure}[ht!]
\plottwo{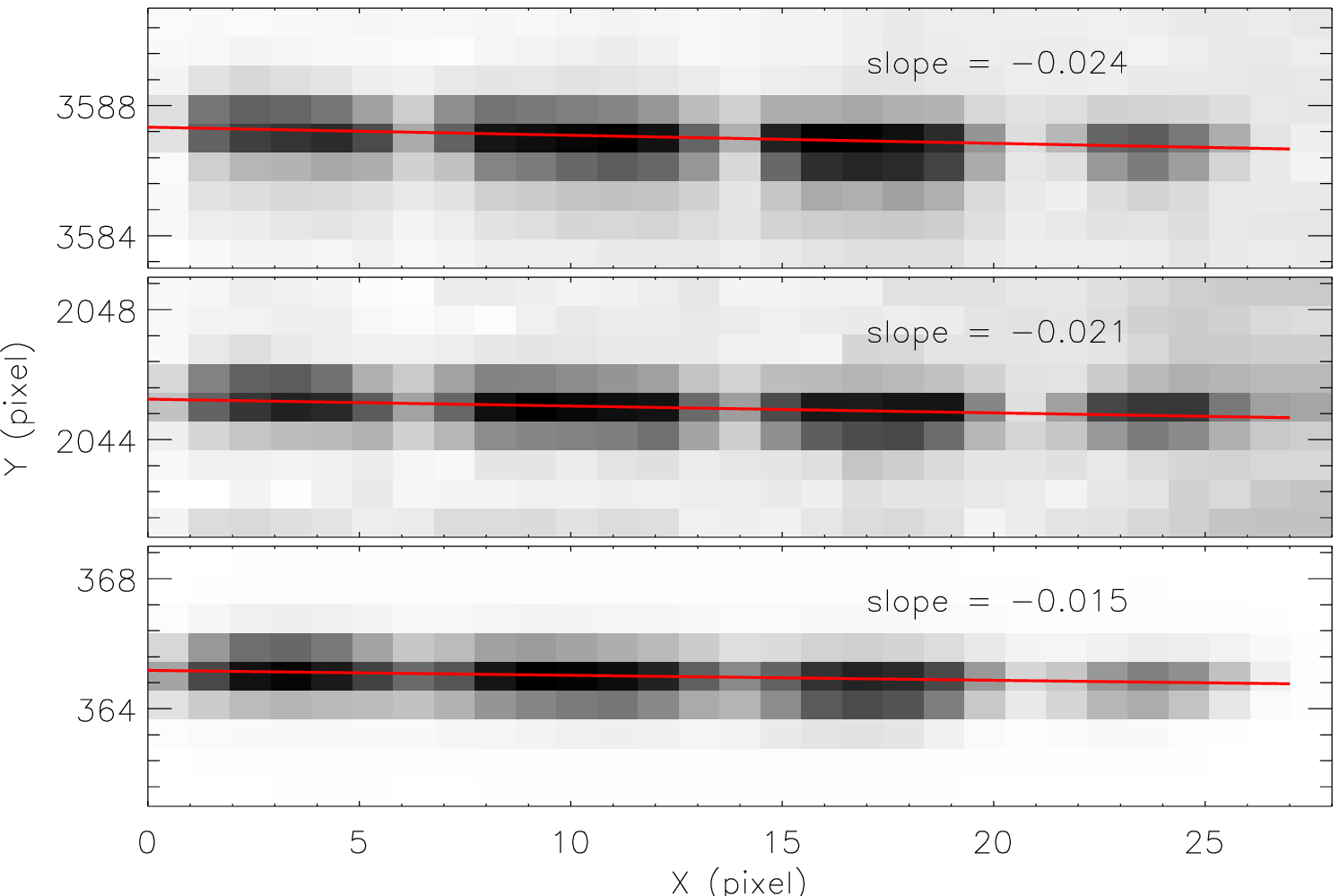}{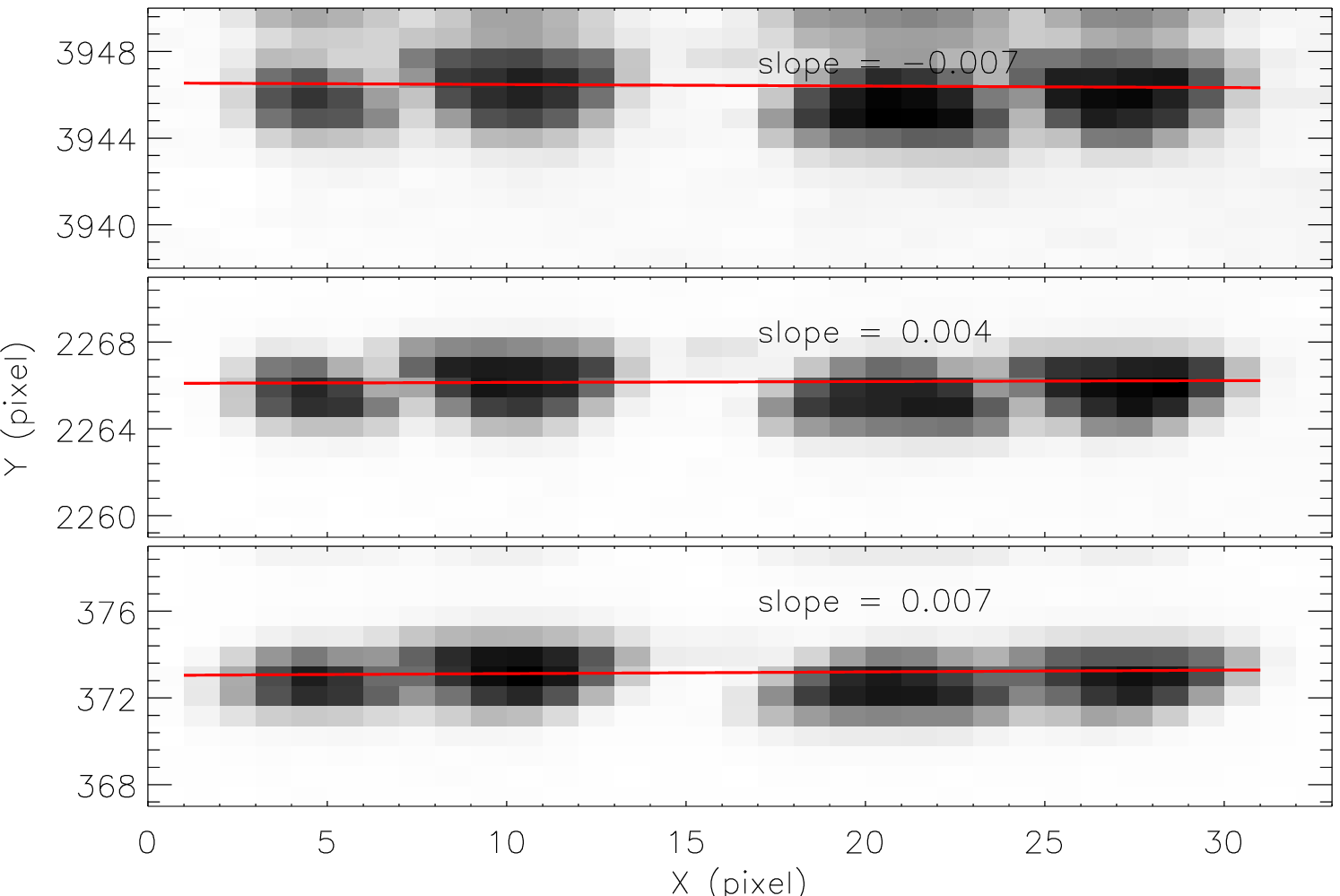}
\caption{Close up view of 3 selected ThAr lines on the 2D frame in both the one-fiber (left) and two-fiber (right) spectral modes. The lines were chosen to be located as far to one another as possible along the order. The red slopes are the results of the linear fit for each line. \label{fig:tilt}}
\end{figure}

\rep{Thanks to the procedure that determines the crude wavelength solution, we now have the position in pixel of many ThAr lines}. The script selects a stamp from the 2-D ThAr master spectrum centered on each identified line with an amplitude higher than 50 ADUs \rep{to avoid small lines, and lower than 65\,000 ADUs to avoid saturated lines}. The stamp is as wide as the order's trace, and as high as 4 times the expected FWHM of the spectral lines. To determine the slit tilt, the script does a cross-correlation of the corresponding line profile from the 1-D arc spectrum with each column in the stamp. This determines the center of light across the spectral line. The tilt is defined as the slope of a robust linear fit of the measured centers. The tilt is determined for the whole order at once (see Figure~\ref{fig:tilt}). Because there is a clear trend in the tilt determined as a function of wavelength (or Y pixel along the chip), a robust linear fit is done to determine the tilt to correct for as a function of Y pixel for each order.

A final wavelength solution is found later \rep{(see Section\,\ref{subsec:identify})}, but for now only the slit tilt value for each order is required.

\subsection{Flatfield normalization}\label{subsec:flatnorm}
\begin{figure}[ht!]
\plottwo{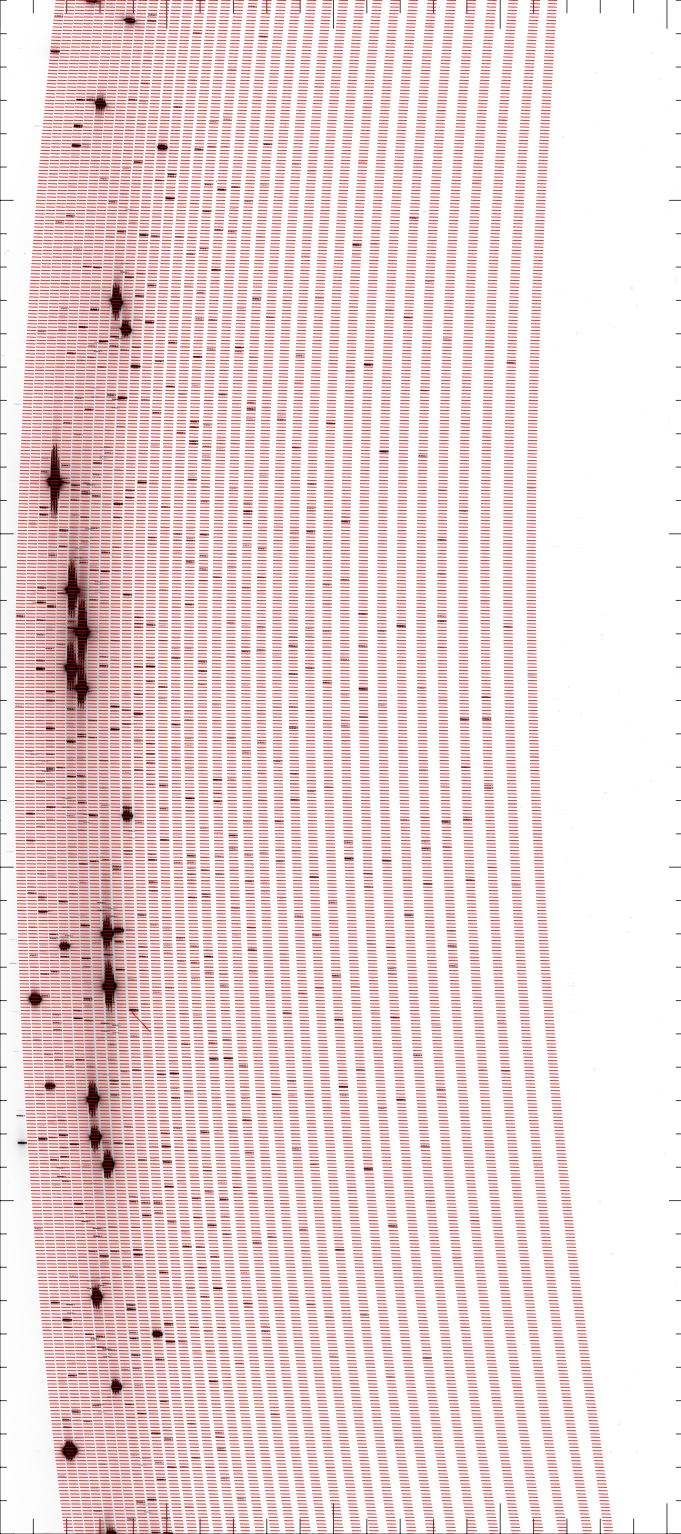}{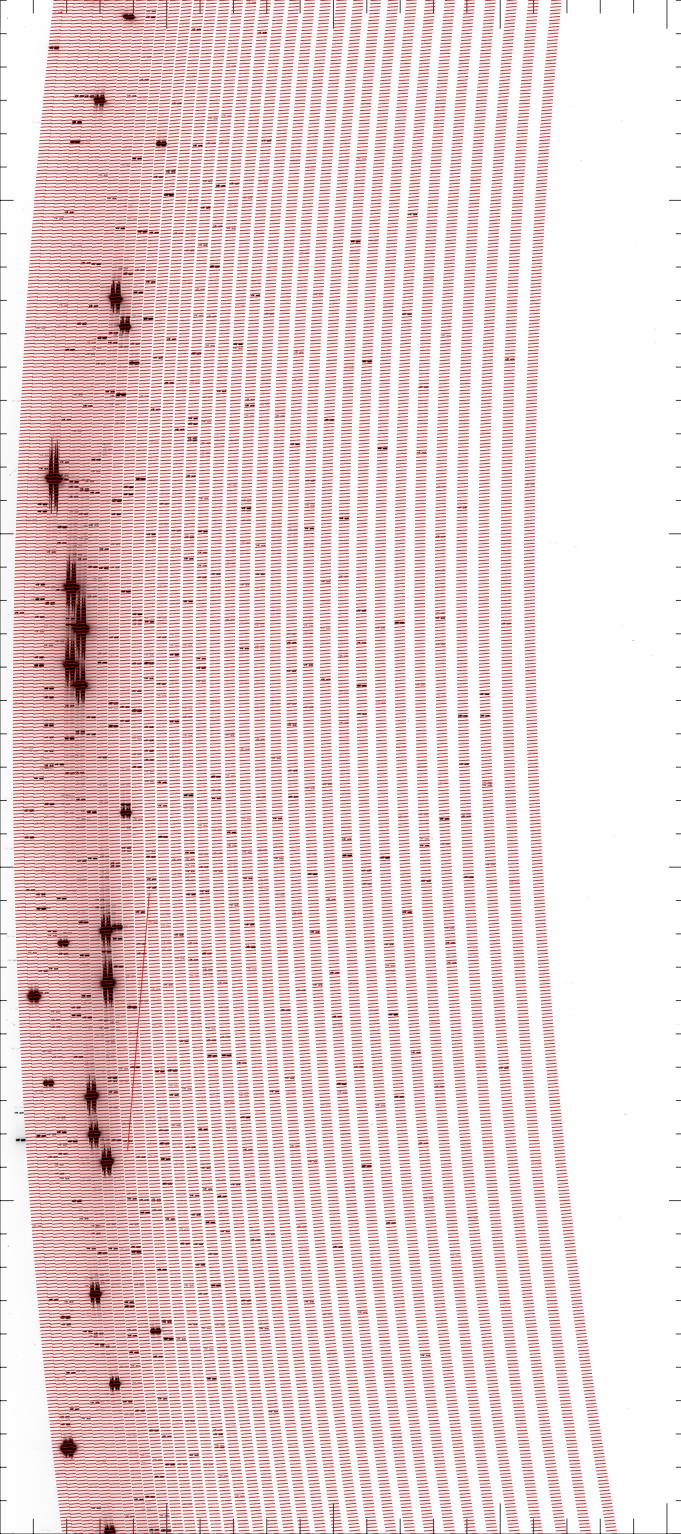}
\caption{Examples of ThAr spectra (one-fiber spectral mode to the left and two-fiber spectral mode to the right) and the orders' traced in red. Each red line has the slope corresponding to the slit tilt at the corresponding wavelength. \label{fig:interpol}}
\end{figure}

\noindent \texttt{Function: reduce} \\
\indent Now that the trace and the slit tilt are known for all the orders (see Figure~\ref{fig:interpol}), the pipeline can normalize the flatfield. Each order (and each fiber independently per order in the two-fiber spectral mode) is rectified and converted into a straight stripe. 

\subsubsection{Orders rectification}\label{subsubsec:rec}

At each line, a bilinear interpolation is used to get the flux within an order that goes along a slope corresponding to the slit tilt, and centered on the order's trace. The interpolated vector becomes the corresponding line of the rectified order. Science frames are later rectified in the exact same way.

\subsubsection{Normalization}\label{subsubsec:norm}

The pipeline fits a polynomial of the order 6 to each of the columns of the rectified flatfield. Each column is divided by its fit. The resulting normalized flatfield frame is a mosaic of all the rectified orders aligned one next to another, and all the values are around 1. The edges of the orders are suffering from some sampling effects caused by the rectification (from the interpolation), and are later excluded from extraction.

\subsection{Final line identification}\label{subsec:identify}% and wavelength solution}
\noindent \texttt{Functions: reduce, extract, find\_lines, wavel\_sol} \\
\indent The final arc master spectrum is rectified as described in \S~\ref{subsubsec:norm}, and extracted using a simple sum of the columns in each order/fiber. There is no clipping or cosmic ray rejection method implemented at this point. The last 300 pixels are rejected, since this region is severely affected by vignetting.
This final extraction is giving the best resolution, since the tilt is corrected before the spectrum is extracted. 
The crude wavelength solution of the new 1D ThAr spectrum is applied again as described in \S~\ref{subsec:tilt}. It is used to find again the spectral lines, which are fitted individually assuming a Gaussian profile. This gives a real positions in pixels for each ThAr line identified from the ThAr line list \citep{atlas}. 
%A polynomial fit of the order 4 of those pixel positions as a function of wavelength is used as the final wavelength solution to the science spectra.

\subsection{Illumination correction}\label{subsubsec:illum}
\begin{figure}[ht!]
\plotone{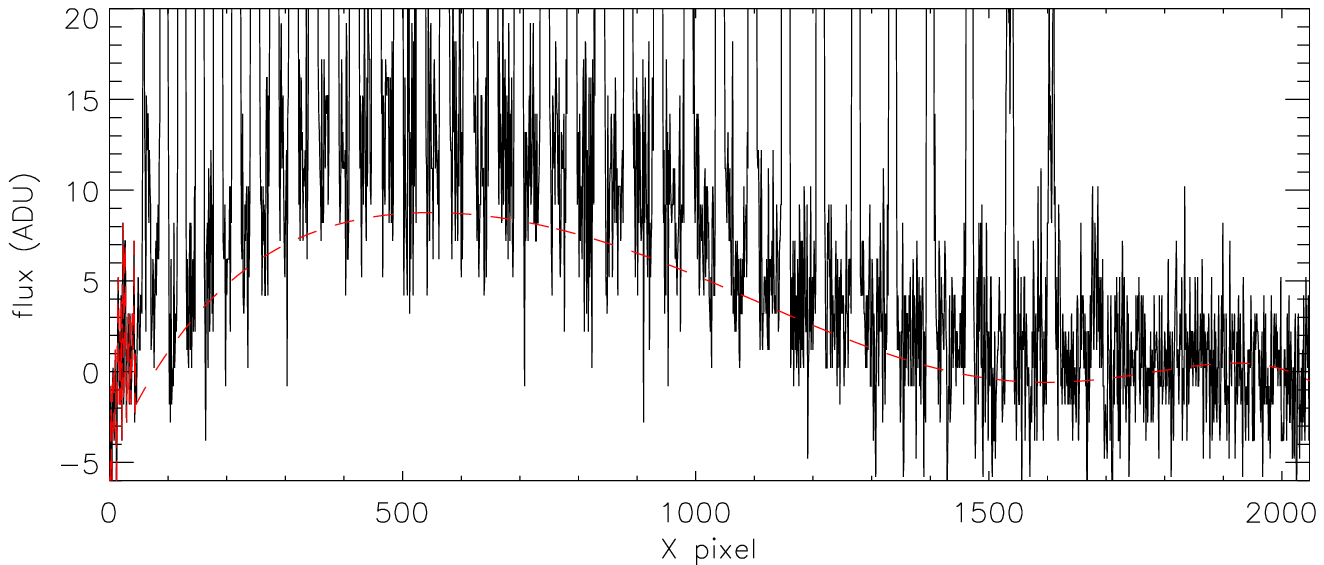}
\plotone{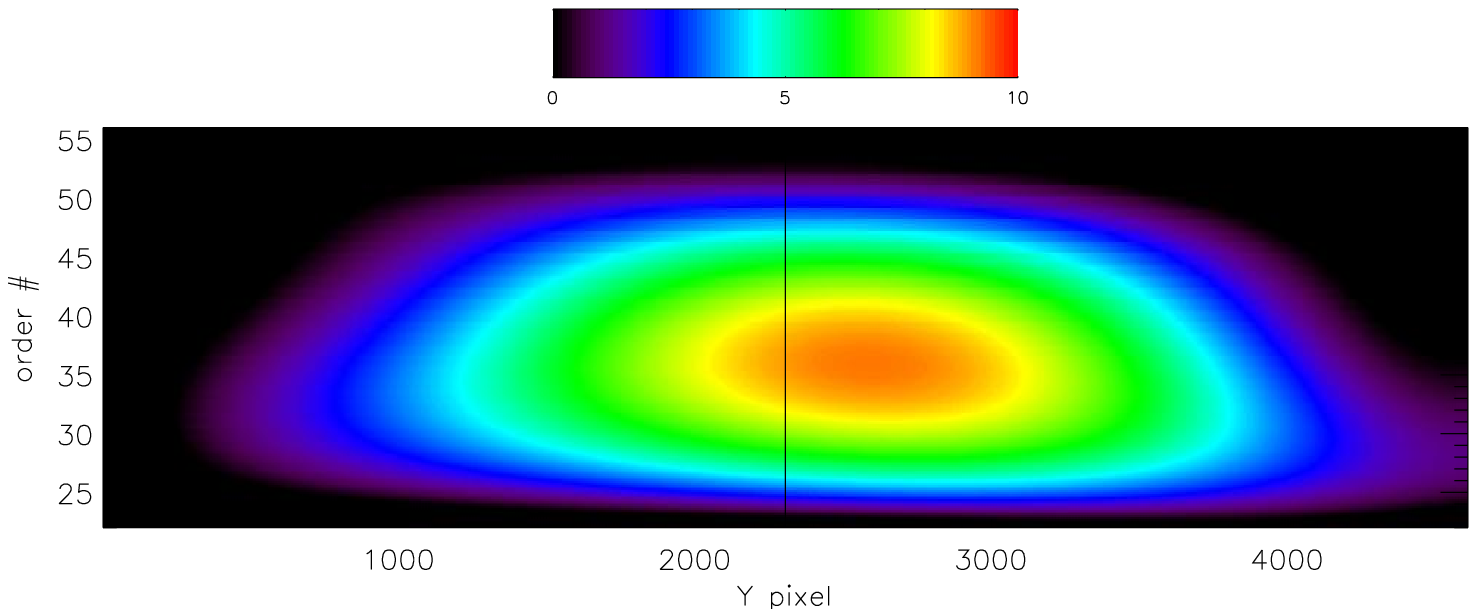}
\caption{{\it Top}: Cut of the 2D flat-field frame across the orders in the middle of the detector. The red-dashed line is the fit to the flux measured in the area between orders. {\it Bottom}: 2D surface of the illumination. The black vertical line marks where the cut plotted on the top panel was extracted. \label{fig:illum}}
\end{figure}

\noindent \texttt{Function: illumcor} \\
\indent The orders are well separated, even on the reddest side of the spectrum (i.e., down to order 22). Indeed, there are at least a few pixels between orders where the flux should be down to $\sim$0 ADU. However, there is flux detected in these regions towards the reddest orders. And a stronger signal in the spectrum gives a stronger effect. To correct for this effect, the pipeline fits and subtract the flux measured in the area between orders on each of the science exposures (unless the user specifically chooses to avoid it) following those steps:
\begin{itemize}
\item Measures on the 2D frame the level of illumination between each order. This is done line by line, and looking at each interval one by one. If the interval is 3 pixel wide or less, the illumination is assumed to be the minimum value. If it is wider, it is assumed to be the minimum of a polynomial fit of order 3 of the interval (to avoid strong dependence on the noise). All the values are recorded in a matrix that contains the background illumination value for each order in each line.
\item Smooths the background illumination values along the Y axis (lines) using a 40 pixels bin.
\item Line by line, fits a polynomial of order 6 to estimate the background illumination values within the orders (see the top panel of Figure~\ref{fig:illum}), and stacks all the results in a 2D image.
\item Rectifies the traces and corrects the slit tilt, just as described in \S~\ref{subsubsec:rec}.
\item For each order, fits a surface using a polynomial of order 5, and creates the final background illumination frame to subtract from each of the reduced science frames (see the bottom panel of Figure~\ref{fig:illum}).
\end{itemize}

Note that bluer wavelengths (around and past X pixel 1500) are left out of the fit, and the background often diverges. It was decided to leave that section out, so the polynomial fit would be less constrained while trying to reproduce the background at redder wavelengths. Also, this causes no real losses, since there is no more useful signal in that region. 

\begin{figure}[ht!]
\plotone{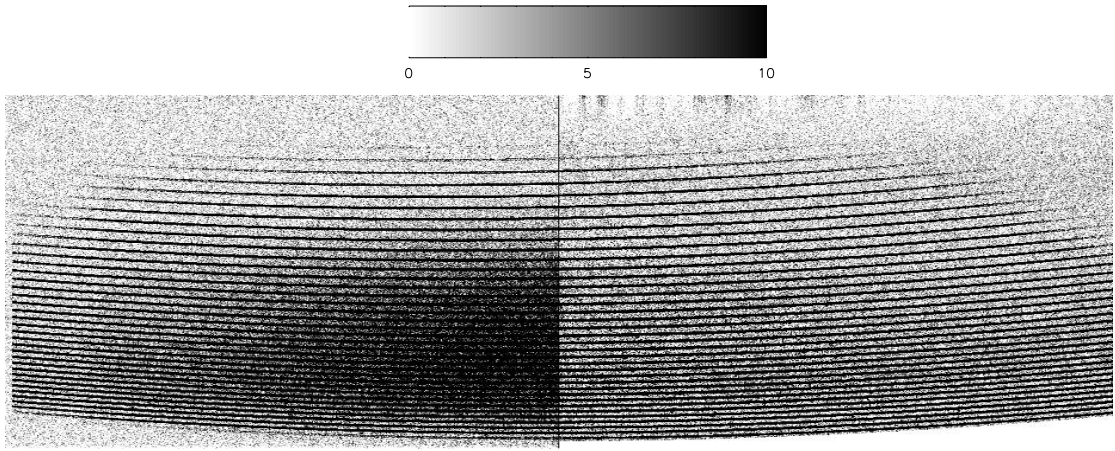}
\caption{Sample 2D frame of the spectrum of BD+28 4211. The left half is the raw frame, while the right half was corrected for contaminating background illumination. The function {\it illumcor} does not correct the 2D spectrum before the orders are rectified, and this figure was created to illustrate the importance of removing the light contamination. \label{fig:illumcor}}
\end{figure}

\rep{The function {\it illumcor} corrects for any contaminating light source, such as stray or scattered light. Such a correction becomes even more important as the exposure time is long. Figure\ref{fig:illumcor} shows an example non-rectified 2D science frame without (the left half) and with (the right half) background illumination correction.} 

%\textcolor{red}{Is incidence the correct word here?}

\subsection{Final extraction}
\noindent \texttt{Functions: overs\_corr, reduce, illumcorr, extract} \\
\indent At this point, the pipeline is ready to reduce the science frames and extract the spectra. 

\subsubsection{Primary reduction}

Each science frame is reduced following those steps:
\begin{itemize}
\item Overscan correction.
\item Bias subtraction.
\item Order rectification and slit tilt correction as described in \S~\ref{subsubsec:rec}.
\item Background illumination correction as described in \S~\ref{subsubsec:illum}
\item Division by the normalized flat obtained as described in \S~\ref{subsec:flatnorm}.
\end{itemize}

\begin{figure}[ht!]
\plotone{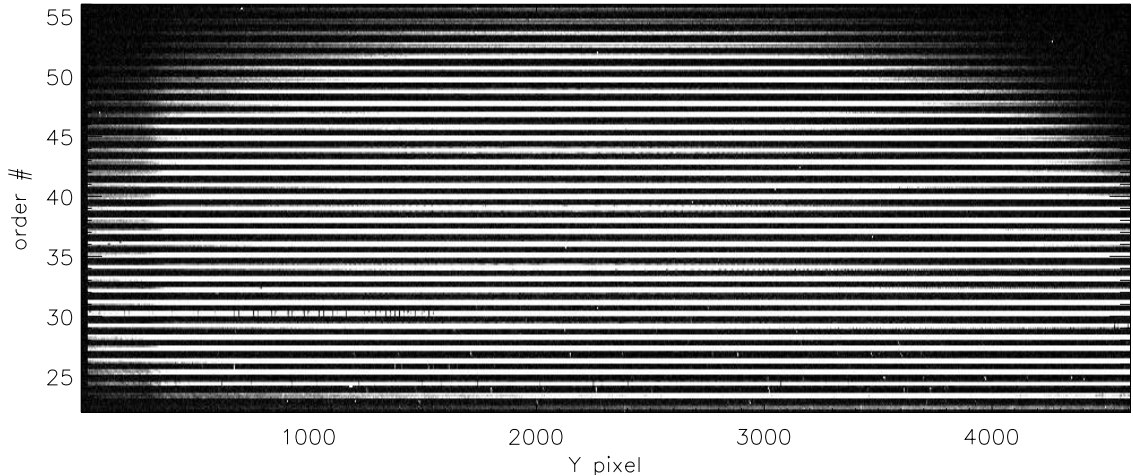}
\caption{Reduced 2D science spectrum. All the orders are now rectified and stacked next to one another. \label{fig:reduce}}
\end{figure}

\subsubsection{Spectral extraction and wavelength calibration}

Each order is extracted using a simple sum of the lines. There are no cosmic ray rejection or any clipping implemented at the moment.

For each extracted order, the pipeline calculates a wavelength solution using a polynomial fit of the order 4 on the pixel position of the ThAr lines (identified as described in \S~\ref{subsec:identify}) as a function of central wavelength. To apply this solution, the extracted spectrum is first linearized by resampling each order using a spline function. The step is chosen to be determined by the highest resolution measured within the order wavelength range.

It is possible to skip the wavelength solution, if the user prefers to do a separate calibration. In this case, the arc spectrum and the science spectra are simply extracted and saved.

The spectra are not normalized by the continuum nor flux calibrated. Spectra of the same target are not combined to a single spectrum, but instead recorded in separate fits files. The wavelength solution is not corrected using telluric lines and is not adjusted to correct for heliocentric motion. In brief, it is a simple sum of the flux. Any correction, combination, calibration or normalization needs to be done separately.

\section{Organization of the reduced data} \label{sec:organization}

The reduced spectra are multi-extension fits files. The filename is the same as the original file, with the prefix ''ext\_''. Each extension has one section of the spectrum, corresponding to one of the orders. For the one-fiber spectral mode, there are 35 extensions, with each order from 22 to 57 on each extension. For the two-fiber spectral mode, there are twice as many extensions, as for each order, there is the sky and the target spectra which are each on a different extension. Each extension contains a unique wavelength solution for the corresponding order, using the keywords CRVAL1, CRPIX1 and CD1\_1.

In the two-fiber spectral mode, since the target and the sky are recorded in separate extensions, one needs to perform the final sky subtraction outside of DRAG{\sc races}. Note that because the sky fiber has a smaller transmission than the target fiber, one has to apply a factor 1.25 to the sky spectrum before subtracting it from the target fiber signal.

%\section{Performances} \label{sec:performance}

\section{Comparison with OPERA} \label{sec:opera}

It is quite surprising, yet reassuring to compare the extracted spectra from OPERA and DRAG{\sc races}, and see how similar they are. Yet, there are some differences, most of which can be explained by some inherent differences between the two pipelines.

The fundamental difference between the two pipelines relies on the method for extracting the fluxes. OPERA uses a tilted rectangular aperture which is oversampled in a sub-pixel grid, where they apply either the sub-pixel sum or the optimal extraction method \citep{horne1986}, while DRAG{\sc races} uses the simple sum of pixels in a rectified row.  The use of optimal extraction explains why OPERA is more robust to bad pixels and cosmic rays and the oversampling explains the slightly higher resolution obtained by OPERA in the on-slice mode. On the other hand the main advantage of DRAG{\sc races} is how straightforward it is to download and run it, without further installation required, and how easy it can be for anyone with an IDL licence to customize it. Table \ref{tab:dragraces_vs_opera} presents a summary of features to compare the two pipelines.

\begin{figure}[ht!]
\plotone{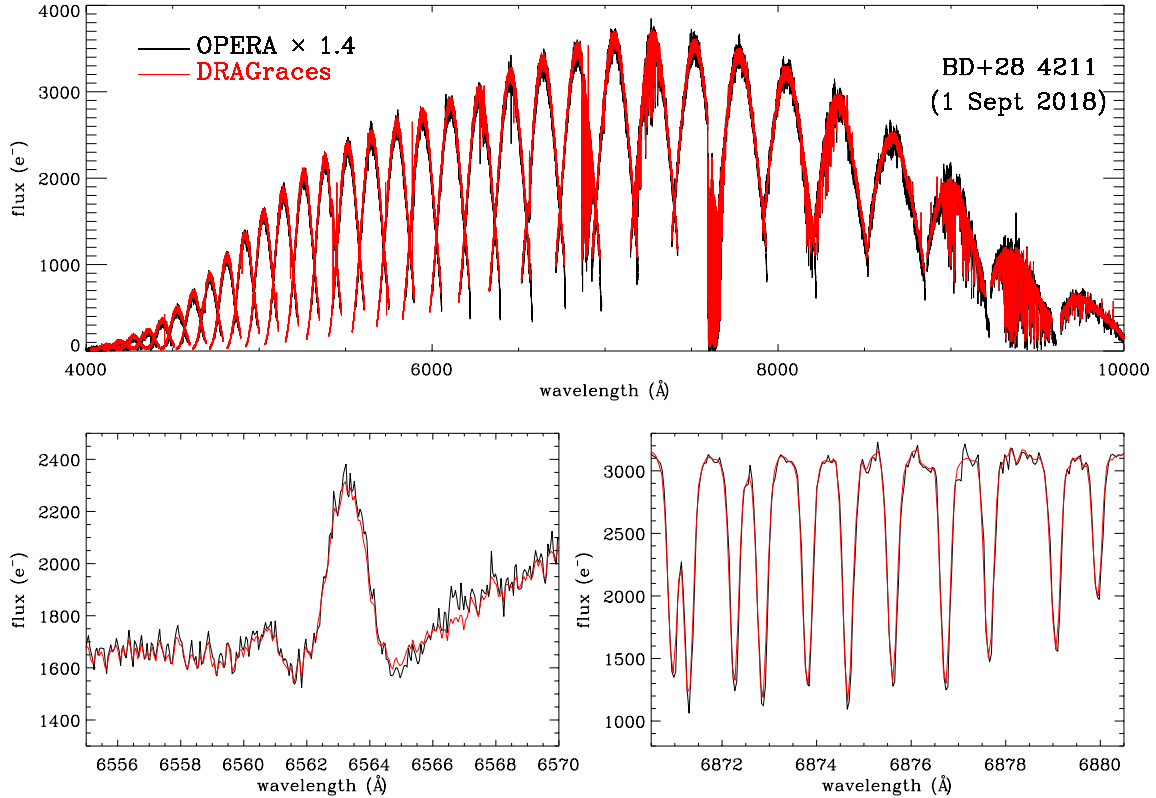}
\caption{Comparison of one spectrum of the star BD+28\,4211 observed on 1 September, 2018, and reduced with both OPERA and DRAG{\sc races}. The observations were taken in two-fiber spectral mode, and the plotted spectrum is that of the star. The spectrum was not flux calibrated, not normalized, not corrected for any instrumental shift in wavelength, nor corrected for the Heliocentric velocity. The spectrum extracted with OPERA was multiplied by a factor 1.4 to match the DRAG{\sc races} extraction. This difference comes from the different wavelength sampling between the two extractions. The top panel show the whole spectrum as extracted. The bottom panels show a close up on the H$\alpha$ line (left) and telluric lines (right). \label{fig:comp}}
\end{figure}

\begin{table}[htbp]
\centering
\caption{Comparison between DRAG{\sc races} and OPERA}
\label{tab:dragraces_vs_opera}
\begin{tabular}{lcc}
\hline
Feature & DRAG{\sc races} & OPERA \\
\hline
Language                     & IDL      & C/C++ and Python or Makefile \\
Open source                  & yes      & yes \\
License type                 & MIT      &  GPL 3.0     \\
Repository for download      & yes\footnote{\url{https://github.com/AndreNicolasChene/DRAGRACES}} &  yes \footnote{\url{https://github.com/CFHT/OPERA}}  \\
Online documentation         & yes\footnote{\url{http://drforum.gemini.edu/topic/graces-pipeline-dragraces/}} &  yes \footnote{\url{http://wiki.lna.br/wiki/espectro}}  \\
Identify files for reduction              & yes      & yes \\
Interactive                  & no       & no  \\
Run automatically            & yes       & yes  \\
Stack calibration data       & yes      & yes  \\
Correct for overscan         & yes      & no  \\
Subtract bias                & yes      & yes  \\
Correct for scattered light  & yes, by global polynomial fit      & no \\
Perform flat-fielding        & yes, on 2D frame  & yes, on 1D spectrum \\
Correct for blaze            & no          & yes   \\
Account for slit tilt        & yes          &   yes     \\
Method to compensate slit tilt & interpolation / rectification    &  use a tilted aperture  \\
Allow variable slit tilt     & yes      & yes, but constant tilt per order \\
Extraction aperture shape    & rectified rectangle (1 x n pixels)   &  oversampled tilted rectangle (m x n pixels)\\
Allow oversampling \footnote{Oversampling meaning sampling spectral elements at steps smaller than 1 pixel}          & no       & yes   \\
Extraction method            & sum      & sum or optimal extraction\footnote{Optimal algorithm by \citet{horne1986}} \\
Calculate flux errors        & no        & yes   \\
Robust to cosmic rays        & no      & yes \\
Wavelength calibration       &  yes     &  yes  \\
Line detection method           & convolution with a comb function & cross-correlation with 2D inst. profile \\
Line measurement method       &  Gaussian fit    &  multiple Gaussian fit  \\
Line rejection method        &  sigma-clipping  & sigma-clipping  \\
Th-Ar atlas                  & \citet{atlas} & \citet{LovisandPepe2007} \\
R=$\lambda / \Delta \lambda$ (1-fiber)    &  $52.8\pm 0.1$ & $56.7\pm 0.7$ \\
R=$\lambda / \Delta \lambda$ (2-fiber)    &  $36.3\pm 0.3$ & $35.7\pm 0.9$ \\
RV accuracy              &    $\sim 500$~m/s      &  $\sim 100$~m/s     \\
Provide RV correction            &  no   &   yes, heliocentric   \\
Continuum normalization & no        & yes   \\
Products in FITS format      & yes       &   yes \\
Development status           & on-going  & interrupted  \\
\hline
\end{tabular}
\end{table}

\subsection{Flux, Signal-to-noise Ratio and Wavelength Solution}

At first look, the spectrum extracted with OPERA spectrum has lower flux. But this is just a result of the re-sampling performed by OPERA to optimize the spectral resolution. The re-sampling gives a wavelength step equivalent to $\sim 0.7$ pixel (the exact value is not public), which explains that the spectrum has to be multiplied by a factor 1.4 to match the flux of the spectrum produced by DRAG{\sc races} (see Figure\,\ref{fig:comp}).

Spectra extracted with OPERA and DRAG{\sc races} have the same signal. A close up on the H$\alpha$ line (Figure\,\ref{fig:comp}) shows that the two extractions give the same spectrum, for the most part. Because the DRAG{\sc races} sampling is more coarse, it has a higher signal-to-noise per wavelength step. Yet, we can recognise the same noise pattern in the two spectra. 

Finally, the wavelength solutions are identical for the two pipelines. A close up to any narrow telluric line shows that there is no apparent shift between the two spectra. This is reassuring, since the wavelength solution seems to be independent to the assumptions used in both codes.

\subsection{Resolution}

\begin{figure}[ht!]
\plotone{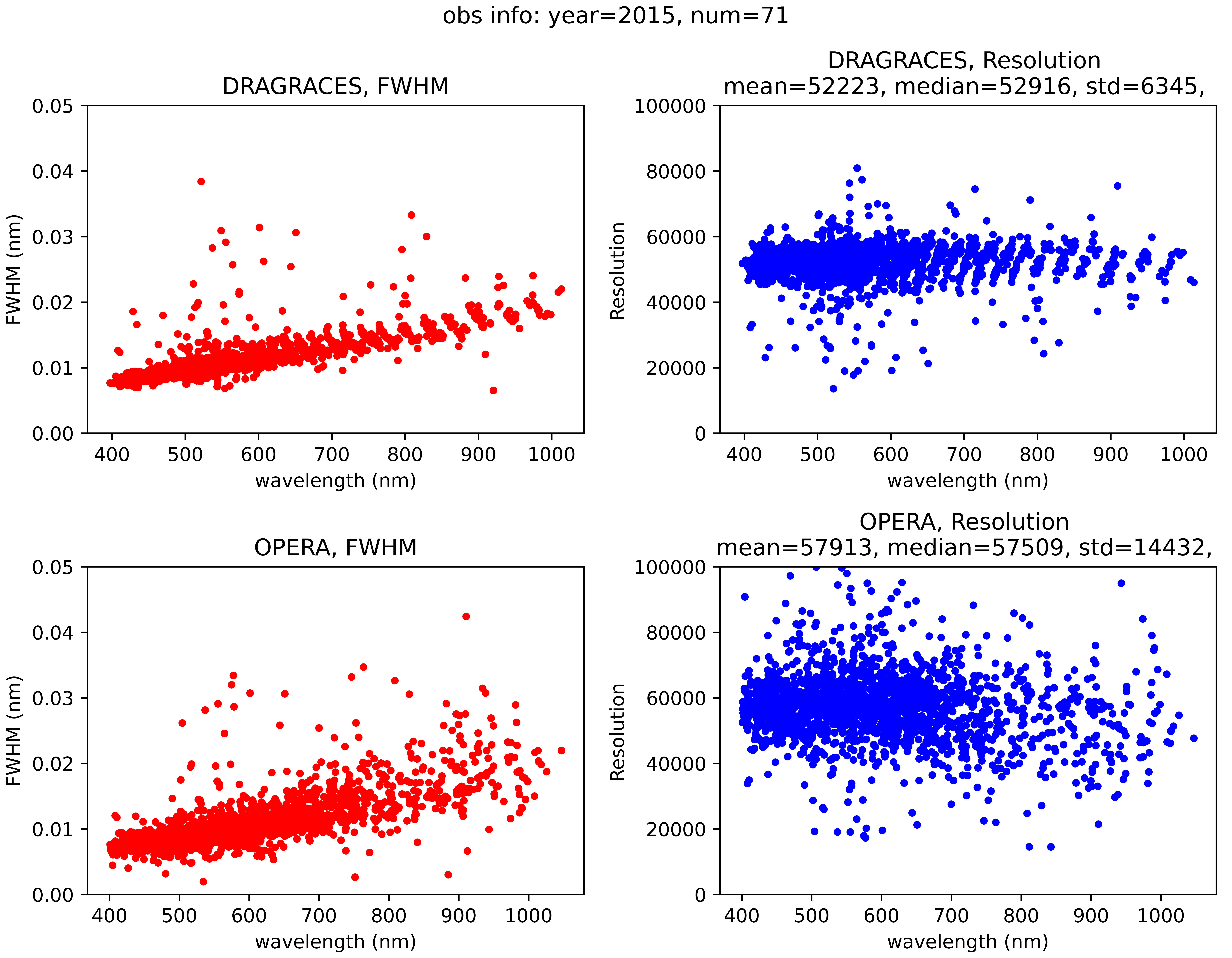}
\caption{FWHM (red, left) and resolution power $R$ (blue, right) of the emission lines selected in a ThAr spectrum extractred using DRAG{\sc races} (up) and OPERA (down). The ThAr spectrum was observed in 7 August, 2015, in the one-fiber spectral mode. \label{fig:resolution1f}}
\end{figure}

\begin{figure}[ht!]
\plotone{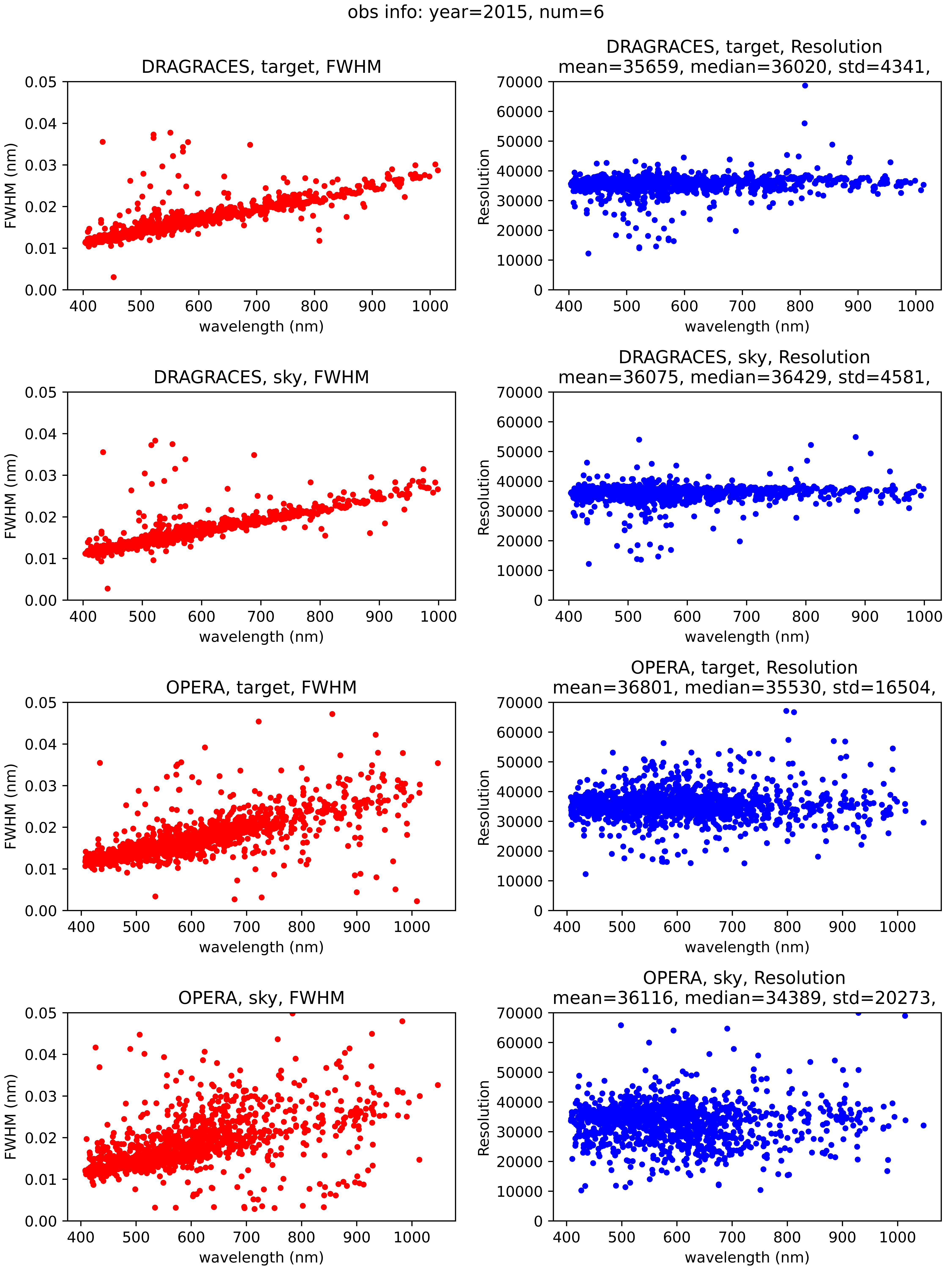}
\caption{Same as \ref{fig:resolution1f}. The ThAr spectrum was observed in 6 June, 2015, in the two-fiber spectral mode. Both the results for the target and the sky fibers are plotted. \label{fig:resolution2f}}
\end{figure}

We extracted a ThAr spectrum using the two pipelines to compare the spectral resolution that they achieve. On the extracted spectra, we selected all the lines with a signal-to-noise ratio superior to 10, rejected lines that are within one resolution element from another, and rejected lines that are not identified in the ThAr line list \citep{atlas}. For each selected line, we used a Monte-Carlo optimiser and a Gaussian model to fit the profile, while trying different wavelength interval cutoff to get the best result. The results are plotted in Figures\,\ref{fig:resolution1f} and \ref{fig:resolution2f}. In the one-fiber spectral mode, the results are coming from the science fiber (usually pointed at the target). In the two-fiber spectral mode, there is one result for each fiber, the one usually pointed at the target, and the one usually pointed at the sky during nighttime observations. In red (left panels) are plotted the FWHM in nanometer (nm) of the selected lines as a function of wavelength. In blue (right panels) are plotted the resolution power $R=\Delta\lambda/\lambda$. 

%\textcolor{red}{ML: For Figures~\ref{fig:resolution1f} and \ref{fig:resolution2f} can we get the axis of the plots to be the same for direct comparisons to be easier?}

\subsubsection{DRAG{\sc races} spectral resolution}
The mean resolution achieved by DRAG{\sc races} in the one-fiber spectral mode is just over $R\sim52.8\pm0.1$\,k. This is below the advertised $R\sim60$\,k for GRACES\footnote{See GRACES webpage \url{www.gemini.edu/GRACES}}. However, this should be expected, since the GRACES specs were determined using the pipeline OPERA, which is optimizing the spectral resolution by re-sampling the resolution elements to a sub-pixel size. The same is not performed by DRAG{\sc races}. In the two-fiber spectral mode, though, the achieved resolution power is $R\sim36.3\pm0.3$\,k, barely missing the advertised $R\sim40$\,k for GRACES. Since the resolution elements in the two-fiber spectral mode are bigger, the need to optimize the spectral resolution is smaller.

One may see that, in the one-fiber spectral mode, even if the resolution remains stable all along the spectrum, it varies linearly within each order. The difference between the blue and the red end is as big as 10\,k. This can be explained by a tilt that was left between the beam and the detector while optimizing the position of the receiver unit in the instrument as a trade off between optimizing throughput and minimizing vignetting. Consequently, the focus is not perfectly uniform along a given order.

%This can be explained by a tilt that was left between the beam and the detector while optimizing the position of the receiver unit in the instrument to optimize throughput and minimize vignetting.

\subsubsection{OPERA spectral resolution}

In the one-fiber spectral mode, OPERA achieves an average resolution that is superior to that of DRAG{\sc races}. The measured resolutions is near $R\sim58$\,k.

\begin{figure}[ht!]
\plotone{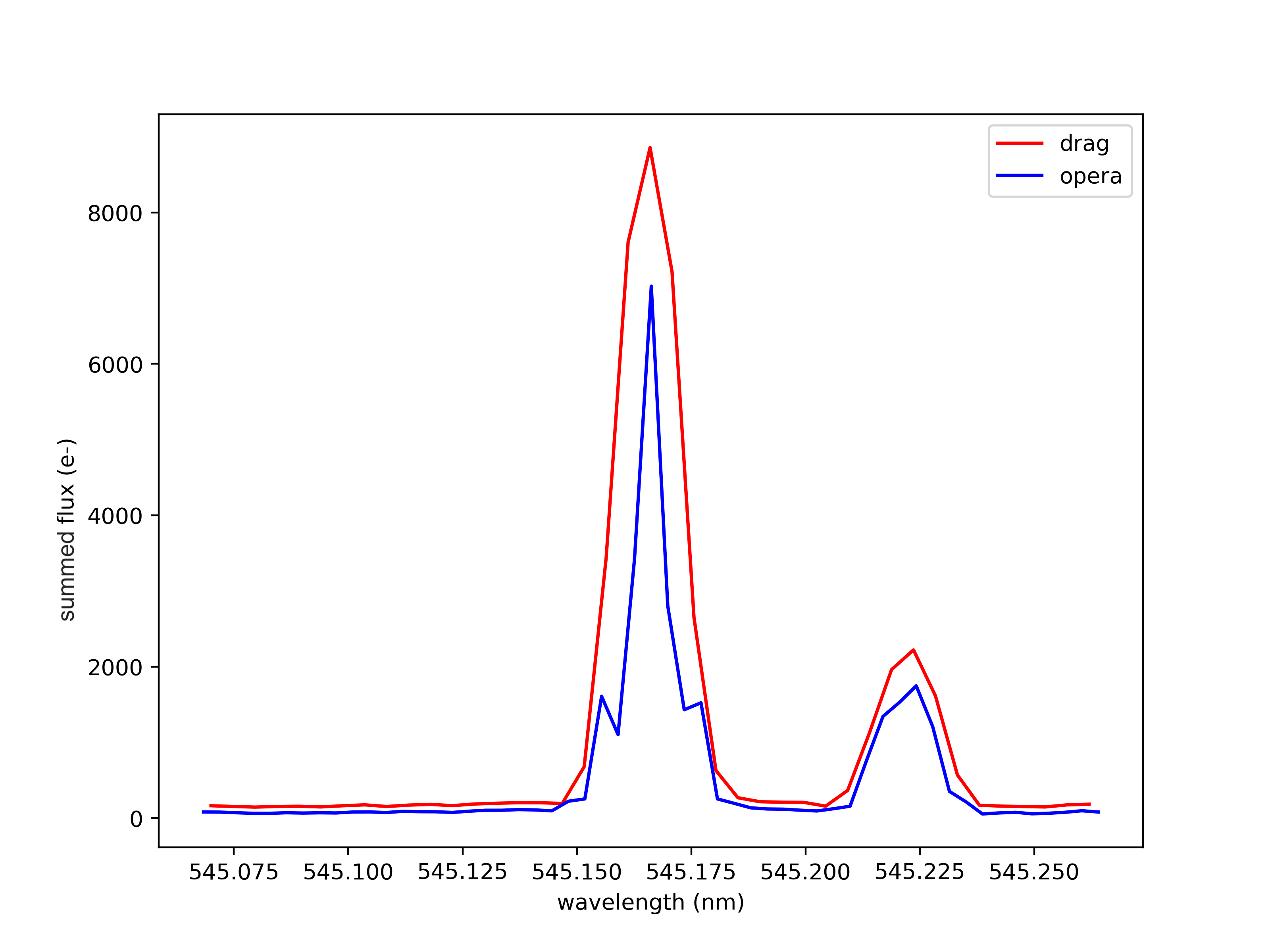}
\caption{Examples of ThAr line profiles obtained using DRAG{\sc races} (red) and OPERA (blue)\rep{ illustrating how the OPERA extraction provided by Gemini alters strong, narrow lines}. For lines with moderate amplitudes, the profiles are identical. For stronger lines, the profile changes from nearly Gaussian to a sharp peak on top of two low shoulders. \rep{It is probably caused by} parameters into OPERA makefile that need to be adjusted in the installation used at the Gemini Observatory. \label{fig:profile}}
\end{figure}

In the two-fiber spectral mode, both OPERA and DRAG{\sc races} deliver a very comparable resolution power ($R\sim36$\,k). 

\subsection{Cosmetics}

For the most part, DRAG{\sc races} and OPERA give very comparable, if not the same results. On the other hand, depending on the case, one can be more affected by cosmetic issues than another. 

From users' feedback, we know that spectra extracted with OPERA can occasionally suffer from strong artificial patterns in the continuum that do not appear in the spectra extracted with DRAG{\sc races}. Those can very likely be corrected by a user who would know how to tune the OPERA parameters, but this requires a fairly high level of expertise with the pipeline. On the other hand, extraction from DRAG{\sc races} is very basic, and can sometimes falls short with some users' expectations. 

One can note the installation of OPERA used at the Gemini Observatory alters the profile of strong and narrow emission lines\rep{, like ThAr or nebular lines}. Indeed, \rep{while narrow lines with amplitudes lower than 6000 e$^-$ (flux as extracted from the chip, not normalized by the exposure time) have the same profile in the OPERA and GRAG{\sc races} extraction, the OPERA extraction of stronger lines gives profiles that diverge from the expected one. Those profiles have two components, i.e., an artificially narrow peak (FWHM$\sim$1 pixel) and a low amplitude plateau} (see Figure\,\ref{fig:profile}). The most likely cause for this \rep{profile alteration} is the choice of parameters used for the extraction. Yet, the fine tuning of the OPERA parameters is beyond the scope for this paper. This profile issue is at the origin of the bigger scatter in the FWHM and $R$ values found in the spectrum extracted using OPERA compared to using DRAG{\sc races} (see Figures\,\ref{fig:resolution1f} and \ref{fig:resolution2f}).

We cannot determine thoroughly in which cases one pipeline is more appropriate than another. This would require extensive testing with a wide variety of data types, which was not performed with either of the two pipelines. This is why it is highly recommended to always compare the two extractions before choosing which pipeline to use.

\section{Conclusions}

DRAG{\sc races} is a pipeline written in IDL to reduce data from the spectrograph GRACES. Its strengths are:
\begin{itemize}
\item it is easy to download and use
\item it requires very low effort to get extracted spectra, as everything is automated, including making file lists
\item it takes only a few minutes to get a spectrum extracted
\item options are straightforward
\item users have the freedom to modify the code to add additional capabilities
\end{itemize}

%\textcolor{red}{ML:The last item is slightly confusing to me.  does just saying `users have the freedom to modify the code to add additional capabilities' cover this?  Do we need to mention github here?}

On the other hand:
\begin{itemize}
\item the reduction and extraction are basic, with no optimization, no cosmic ray rejection, no weighting applied
\item{\rep there are many steps involving interpolation, which causing errors to be further correlated}
\item{\rep there are no error array provided (for the moment)}
\item flux calibration, normalization and/or wavelength solution corrections have to be performed outside of DRAG{\sc races}
\item one needs an IDL licence to run it
\end{itemize}

Future development for improving those weaker points may be planned, assuming GRACES will still be in use for a longer period of time.

Neither OPERA or DRAG{\sc races} underwent any sort of thorough testing. That can be a concern, since the quality of the results is not well determined. On the other hand, considering that both pipelines have been developed completely independently, that they are based on different principles, and that they produce very similar results is in many ways very reassuring. Each of those two pipelines have their advantages and disadvantages, and one should check both before choosing which one they will base their research on. Yet, regardless of that choice, we are now confident that the extracted spectra are sensible, and the science results are not going to be significantly affected by the way the data were reduced.

\acknowledgments

Supported by the Supported by the international Gemini Observatory, a program of NSF's NOIRLab, which is managed by the Association of Universities for Research in Astronomy (AURA) under a cooperative agreement with the National Science Foundation, on behalf of the Gemini partnership of Argentina, Brazil, Canada, Chile, the Republic of Korea, and the United States of America. JLC acknowledges support from NSF grant AST-1816196.


\vspace{5mm}
\facilities{Gemini}


\software{astrolib \citep{La93},  
          imdisp (written by Liam E. Gumley)
          }

\end{document}